\documentclass[prb, amsmath, amssymb, preprint]{revtex4}
\usepackage{epsfig}
\usepackage{graphicx}

\newcommand{\beq} {\begin{equation}}
\newcommand{\eeq} {\end{equation}}
\newcommand{\beqa} {\begin{eqnarray}}
\newcommand{\eeqa} {\end{eqnarray}}

\begin{document}
\title{Numerical Investigation of the Entropy Crisis in Model Glass Formers}
\author{Yisroel Brumer and David R. Reichman}
\affiliation{Department of Chemistry and Chemical Biology, Harvard University
12 Oxford St., Cambridge, MA, 02138}
\date{\today}
\begin{abstract}
We investigate numerically the low temperature equilibration of glassy
systems via non-local Monte Carlo methods. We re-examine several systems that have been studied previously and investigate new systems in order to test the performance of such methods near the putative Kauzmann temperature, $T_K$, where
the configurational entropy is presumed to vanish. Our results suggest
that previous numerical claims in favor of and against a thermodynamic 
transition at a finite $T_K$ must be re-evaluated. Our work provides some
guidelines and suggestions for future numerical investigations of 
disordered systems at high densities and low temperatures.
\end{abstract}
\maketitle

\vskip .2truein
\section{Introduction}
\vskip .2truein

The nature of the glass transition is one of the great unsolved
problems in condensed matter physics.  The kinetic glass transition
(occurring at temperature $T_g$) is phenomenologically defined to occur
when a liquid reaches a threshold viscosity \cite{Angell}. It has long been
speculated that a thermodynamic transition to a disordered solid would
occur if the liquid could be cooled infinitely slowly without
crystallizing.  Kauzmann \cite{Kauzmann} argued that extrapolating the
supercooled liquid entropy into the glassy regime implies that a
temperature (known as the ``Kauzmann temperature'', $ T_K $) exists
where the configurational entropy of the liquid becomes less than that
of the crystal.  Gibbs and coworkers \cite{Gibbs65} resolved this crisis by suggesting that a
phase transition to an ``ideal glass'' occurs at $ T_K $.  Many
theories of glassy behavior contain thermodynamic transitions to an
ideal glass state.  Gibbs and Di Marzio \cite{GibbsDimarzio} postulated that 
a second order
transition occurs at $T_{K}$ while theories based upon mean-field
spin glass ideas contain a random first order transition at $T_{K}$ 
\cite{XiaKirkKirk2b, ParisiThermoParisiThermoLJ, ParisiThermoSoft}.
Stillinger has argued that localized defects must destroy a second
order ideal glass transition \cite{Stillinger}.  In general, localized
excitations (non-mean-field effects) should round any transition that
would occur at $T_{K}$, and recent arguments have been put forward to
access the conditions under which the signatures of mean-field
behavior survive \cite{Eastwood}.  While the above theories in some sense {\em
presuppose} the existence of a transition at $T_{K}$ that drives the
formation of a disordered solid, other theoretical models of glassy
behavior completely avoid the notion of an underlying thermodynamic
transition.  In particular, a large class of kinetic facilitated
models successfully explain many of the features exhibited by
supercooled liquids without invoking the notion of configurational
entropy \cite{Fredrickson1Fredrickson2Butler1JackleRitort}. 
The difficulty of directly probing slowly cooled, low
temperature glassy configurations, either experimentally or
computationally, has made resolution of the issue of the entropy
crisis extremely difficult.

Advances in simulation techniques have made the low temperature
equilibration of certain model systems possible. These techniques
include parallel tempering \cite{Marinari92Falcioni99Neirotti00}, 
histogram re-weighting \cite{Yan} and the use of
specific non-local MC moves \cite{Gazzillo, Krauth, Swapping}. Recent studies of model systems have
produced seemingly conflicting results regarding the possible
existence of an entropy crisis. Grigera and Parisi have demonstrated
that a soft-sphere mixture displays a distinct signature of a
thermodynamic transition near the Kauzmann temperature predicted by
the random first order phase transition scenario \cite{Swapping},
while Santen and Krauth have shown that a two dimensional hard-sphere
system displays no such signature \cite{Krauth}.  It has been
difficult to obtain further insight into these phenomena, as the
simulation methodologies are highly system specific. Although
extremely useful for several systems \cite{Krauth, YukawaSpheres,
KrauthSanten, MoreKrauth}, the cluster algorithm of Krauth
\cite{Cluster} has been hard to generalize to more complex and three
dimensional systems, while the swapping method used by Grigera and
Parisi fails for systems with attractive interactions.  Other methods,
such as simulated annealing and parallel tempering, have been found to
equilibrate slowly in the glassy regime\cite{Michele}.

In this paper, we re-examine claims in favor and against the existence
of an entropy crisis at a temperature $T_{K}$. We carefully explore the nature of the thermodynamic
signature in the heat capacity of the soft-sphere system studied by
Grigera and Parisi and study thermalized
versions of system studied by Santen and Krauth in both two and three
dimensions. 
A critical assessment is made concerning the
ability of various non-local Monte Carlo techniques to equilibrate
liquids under conditions necessary to probe the putative entropy
crisis. {\em Our conclusions suggest that strong claims for numerical
evidence both in favor and against an entropy crisis in all systems
studied to date must be re-evaluated.}

\vskip .2truein
\section{Computational Details}
\vskip .2truein

When a liquid is deeply supercooled, its microscopic motion is
characterized by rapidly growing relaxation times. This slow
relaxation renders the use of conventional molecular dynamics and
Monte Carlo techniques impractical for obtaining well equilibrated
configurations near the glass transition temperature.  Recently,
several methodologies have been introduced in an effort to overcome
the problems associated with the rapid onset of sluggish dynamics that
hinder equilibration with local simulation techniques.  Parallel
tempering methods have been shown to speed up equilibration above the
mode-coupling temperature $T_{c}$ \cite{YamaKob}, but have not been useful in
obtaining well equilibrated configurations significantly below this temperature 
\cite{Michele}.  Other
techniques, such as expanded ensemble methods \cite{PabloYan}, multi-canonical
algorithms \cite{Berg}, and the density-of-states Monte Carlo approach 
\cite{WangWang2, Yan2Pablo2} may provide
routes to obtaining equilibrated configurations at low temperatures,
although few systematic studies have been performed for glassy systems
with these techniques and we will not discuss them further here
\cite{Faller, Note}.

Conventional Monte Carlo techniques augmented with non-local moves
offer a powerful approach to overcoming ergodicity problems associated
with rough free energy landscapes.  These methods allow for moves that
interchange particles or groups of particles while still maintaining
detailed balance.  Even if such moves are rarely accepted, they
greatly accelerate sampling by providing routes for relaxation ({\em
e.g.} the hopping of a particle out of the cage comprised of its
neighbors) that would take many local Monte Carlo moves to achieve.
Recently, two types of non-local Monte Carlo methods have been applied
to the study of glassy systems: The cluster pivot algorithm (CPA) of
Krauth and coworkers \cite{Krauth} and the swap Monte Carlo (SMC)
algorithm applied by Grigera and Parisi to
the study of the entropy crisis \cite{Swapping}.

In this work we introduce a trivial modification of the SMC approach,
called swap-sector Monte Carlo (SSMC), that is optimized for 
equilibrating polydisperse systems. As with SMC, the positions of
particles of different sizes are swapped at varying intervals. Noting
that the probability of swapping falls off quickly as the ratio of
particle sizes differs from 1, we introduce a swap-sector, where
particles are only swapped with those that differ in radius by less
than a tunable parameter $ \Delta \sigma $. In the limit of large $
\Delta \sigma $, standard SMC is recovered.  $ \Delta \sigma $ can be optimized for
efficiency, as a large $ \Delta \sigma $ yields a low acceptance
ratio, and a small $ \Delta \sigma $ allows pairs of particles to be
swapped back and forth repeatedly. This method has been used to
equilibrate both hard and soft-sphere polydisperse systems, facilitating
direct comparison between the two.
 
Three different systems are simulated in this work. The first is 
a binary soft-sphere system identical to that
studied by Grigera and Parisi, with \beq V = \sum_{i \neq j}^N
\left(\frac{\sigma_i + \sigma_j}{|\bf{r_i} - \bf{r_j}|}\right) ^{12},
\eeq where $ \sigma_1 / \sigma_2 = 1.2 $, $ \sigma_{12} =
\frac{1}{2}(\sigma_1 + \sigma_2) $ and $ \epsilon_1 = \epsilon_2 = 1
$. The units are set by choosing $\sigma_1 = 1$. Runs are executed by
varying the temperature with constant volume, chosen so that $\rho^* =
\rho\sigma_x^3 = 1$, where \beq \sigma_x^3 = \frac{1}{4}\sigma_1^3 +
\frac{1}{2}\sigma_{12} + \frac{1}{4}\sigma_2^3 \eeq is the effective
radius defined by conformal solution theory \cite{HansenMcDonald}.
All simulations in this work employ periodic boundary conditions. 
Systems of 34, 60, 70 and 258 particles are studied, 
consistent with previous studies. Although clearly too small to study dynamics
in a regime where the {\em dynamical} correlation length is noticeably 
growing, our studies should provide a useful probe of the underlying 
thermodynamic behavior.

The second system is the
two dimensional hard-sphere system of Santen and Krauth \cite{Krauth},
run in the NPT ensemble. The 256 particles have radii ranging from $
R_1 = 1 $ to $ R_{256} = 20 $, with $ R_i - R_{i - 1} = \Delta $ for
constant $ \Delta $.  Isothermal compressibilities are calculated both
by finite difference of the volume with respect to pressure,
$\kappa_{diff} = -\frac{1} {\langle V \rangle} \frac{\partial \langle
V \rangle}{\partial P}$, and by fluctuations of the volume,
$\kappa_{fluc} = 
\frac{\langle V^2 \rangle - \langle V
\rangle^2}{\langle V \rangle} $.  If the system is well equilibrated
and the data well averaged, $ \kappa_{diff} = \kappa_{fluc} $. This
provides a necessary (but not sufficient) test for equilibration that,
together with convergence and stability of thermodynamic averages, is
applied throughout. 
We also applied the Flyvberg-Petersen method \cite{Flyvberg} to
determine accurate estimates of the statistical error, which
were well below 0.1\% of the heat capacity, even at extremely low temperatures.

Lastly, a polydisperse soft-sphere system is studied with $ \sigma_i =
1 + \Delta*i $ where $ 1 \leq i \leq N $ and $ \Delta $ is chosen so that $R_{max} = 20, $
providing a thermal version of the Santen and Krauth system. Runs are
executed by both varying the temperature with constant volume, and
varying the volume with constant temperature to confirm the validity
of the results (excess thermodynamic properties in soft-sphere systems
depend only on the parameter
$\Gamma= \rho T^{-\frac{D}{12}}$, where $D$ is the dimensionality of
the system). Constant volume is chosen so that $ \rho_{poly} =
\frac{\frac{4}{3}\sum_{i = 1}^{N}\pi r^3}{L^2} = 1 $.  For all of the systems, up
to $ 1 \times 10^8 $ steps of equilibration and $ 1 \times 10^9 $
steps of data collection are used, providing well converged data.
Data accuracy and convergence properties were established through
numerous shorter runs.

In several cases presented below it is useful to have an estimate of
the location of the putative Kauzmann temperature as calculated
through the random first order theory.  Here we follow Parisi and
coworkers \cite{ParisiThermoParisiThermoLJ, ParisiThermoSoft}.  Harmonic solid entropies are calculated for $ N $
particles in $ D $ dimensions as
\begin{equation}
S^{(a)}_{sol} = \frac{ND}{2}\left( 1 + log \left(\frac{2\pi}{\beta}
\right) - \left< \frac{1}{N_{pos}} \sum_{i =
1}^{N_{pos}}log(|\lambda_i|)\right> \right),
\end{equation}
where $ N_{pos} $ denotes the number of positive eigenvalues $
\lambda_i $ of the instantaneous Hessian.  These are averaged over
100-200 different configurations for each temperature, each separated
by $1 \times 10^4 $ steps, which is more than enough to obtain good
statistics.  Liquid entropies are calculated by thermodynamic
integration as
\begin{equation}
S_{liq} = S_{liq}^0 + \beta E_{liq}(\beta) - \int_0^\beta d\beta' E_{liq}
(\beta'),
\end{equation}
where $ S^0_{liq} $ represents the ideal gas entropy in the $ \beta
\rightarrow 0 $ limit and $ E_{liq}(\beta) $ is the average energy at
inverse temperature $\beta$.  Liquid entropies are extrapolated into
the glassy regime by fitting the data as $ S_{liq}(T) = aT^{-2/5} + b
$, a well known fitting form for the entropy of simple liquids
\cite{Yakov, ParisiThermoSoft}.  Inherent structures, when needed, are found by
steepest descent quenches for the binary soft-sphere system, and a
combination of steepest descent and conjugate gradient for the
polydisperse system. Data is averaged over thousands of such
configurations for each temperature. 

\vskip .2truein
\section{Binary Soft-Sphere System}
\vskip .2truein

The first system that we investigate is the binary soft-sphere liquid
defined in Sec. 2.  To demonstrate the robustness of estimates of
the Kauzmann temperature in this system, we calculate $T_{K}$ via the
random first order theory of Mezard and Parisi \cite{ParisiThermoParisiThermoLJ, ParisiThermoSoft} and the inherent
structure-based method of Buchner and Heuer \cite{Buchner1,
Gaussian}.  The approach of Buchner and Heuer is similar to
that of Sciortino, Kob and Tartaglia \cite{Kob1, Kob2}, and yields nearly the same
estimate of $T_{K}$.  Using the approach of Mezard and Parisi and the
definitions of the entropy of the liquid and disordered solid
discussed in Sec. 2, we find $ \Gamma_K = 1.72 $
($\Gamma=\rho T^{-\frac{D}{12}}$) while the method of Buchner and
Heuer yields $ \Gamma_K = 1.69 $.  A previous estimate using the
random first order approach by Coluzzi {\it et. al.} \cite{ParisiThermoSoft}
yielded $ \Gamma_K = 1.65$ analytically and $\Gamma_K = 1.75$ from simulation data.  It is unclear why our estimate differs from that of Coluzzi
{\it et.al.}, but it is most likely due to slight differences in numerical 
implementation.  Regardless of these
differences, a consistent estimate of $\Gamma_{K} \approx 1.7$ emerges
from these approaches.

Recently, Grigera and Parisi implemented the SMC procedure for the
soft-sphere mixture.  While for large system sizes they could not
equilibrate the system at low temperatures (large $\Gamma$), they
found a peak in the specific heat of a small (34 particle) system at
$\Gamma \approx 1.7$.  We have reproduced this calculation in Fig. 1a.
The agreement between the values of the specific heat calculated from
fluctuations and derivatives of the average energy suggest that the
equilibration of configurations has been achieved.  In the inset of
Fig. 1, we display an expanded view of the specific heat for the
34 particle system that includes higher temperatures (lower $\Gamma$
values).  A second interesting signature in the specific heat appears
near $\Gamma=1.25$.  This signature becomes sharper for larger systems,
as shown in Fig. 1B. The sharpness of this peak, together with its
system-size dependence, strongly support the view that this feature is 
thermodynamic in nature.
It is interesting to note that $ \Gamma = 1.25 $
is close to the onset of supercooling at $\Gamma_0 \approx 1.3 $
\cite{BrumerPrePrint}. For systems that are larger still ($N>70$)
the signature near $\Gamma=1.25$ vanishes, as does the ability to
equilibrate the system with the SMC method for $\Gamma > 1.5$ for reasonable 
trajectory lengths.

The appearance of secondary features in thermodynamic quantities
suggests that perhaps the specific heat peak near $\Gamma=1.7$ is not
the result of the type of entropy crisis envisioned in the random
first order theory of glassy thermodynamics.  To investigate this
further, we examine the structures obtained from SMC in small systems
that display a specific heat peak near $\Gamma=1.7$.  In Fig. 2 we
show representative configurations of the system obtained by SMC at
$\Gamma = 1.3$ and $1.8$.  The system is clearly not an amorphous solid, 
but a phase
separated crystal with substitutional defects, and, at very low temperatures,
the phase separated crystal itself.  

The ground state structure found by SMC is a phase-separated crystal.
By examining the inherent structure configurations sampled at various
values of $\Gamma$, {\em we conclude that the peak in the specific
heat results from a gap in the density of states between the set of
all defective crystal configurations and the ground state of the
system.}  As the system size increases, defective crystalline
configurations become harder to locate relative to the set of
amorphous configurations, and the sharp features in the specific heat
vanish, as does the ability of SMC to equilibrate the system
significantly below the estimated value of $\Gamma_{c}$, the location
of the mode-coupling temperature (density)\cite{Swapping2}.  In Fig. 3 we plot the
specific heat for 70 particles obtained from the SMC method.  Note
that the agreement between the heat capacity calculated via
fluctuations and by direct differentiation of the average energy
diverge from each other near $\Gamma_c \approx 1.45$. 


We thus conclude that the specific heat peak found in the binary
soft-sphere system is not a direct consequence of an ideal glass
transition.  One may speculate that the gap between defective
crystalline states and the ground state occurs near $\Gamma_{K}=1.7 $
because the number of amorphous configurations at these energy values
(nearly) vanish, and thus the peak at $\Gamma_{K}=1.7$ is indeed an
indirect consequence of the vanishing of the configurational entropy
associated with glassy states.  The fact that slightly larger systems
(i.e. $N > 70$) than those for which converged specific heats may be
obtained cannot be equilibrated significantly below $\Gamma_{c} \ll
\Gamma_{K}$ makes this reasoning suspect. 
It still remains to be determined if the fact that the predicted Kauzmann 
value  $ \Gamma_c \approx 1.7 $ coincides with the gap between defective crystal
states and the ground state is purely accidental.
 Furthermore, we find that,
while SMC is more efficient than methods like parallel tempering for
$\Gamma \approx \Gamma_{c}$, neither approach is able to equilibrate
large systems significantly below the mode-coupling temperature
(density) in this system.

\vskip .2truein
\section{Polydisperse Systems}
\vskip .2truein

Having seen that states with crystalline order interfere with a direct
investigation of the entropy crisis in the binary soft-sphere system,
we now turn to the analysis of polydisperse systems.  The first goal
will be to re-investigate the completely polydisperse two
dimensional hard-sphere system studied by Santen and Krauth\cite{Krauth}.  Here
we will examine systems in three dimensions, as well as systems at
finite temperatures to address concerns that might arise with Santen
and Krauth's particular system.

First, we calculate the compressibility of a 256 particle two
dimensional hard-sphere system via the SSMC approach.  In Fig. 4 we
show $\kappa^{-1}$ vs. $\rho$ for this system.  Our results agree with
those of Santen and Krauth, although we were able to equilibrate the
system at slightly higher densities.  Interestingly, we have found
that our implementation of SSMC is more efficient than the cluster
algorithm of Santen and Krauth, although this may indeed be a result
of our own inefficient optimization of their technique.  Regardless,
the similarity in the convergence of the two methods is significant,
as the SSMC approach works in arbitrary dimensions, while the method
of Santen and Krauth is difficult to generalize, particularly
to systems with $D>2$ \cite{KrauthPersonal}. By 
fitting the diffusion constant of the larger discs to a power law form,
$D(\rho) \approx (\rho-\rho_{c})^{\gamma},$ Santen and Krauth extract
a ``glass transition'' density of $\rho_{c} \approx 0.805$.  
These diffusion constants are measured by short-time local Monte Carlo
dynamics, which has been shown to accurately yield relaxation times
in two dimensional systems \cite{Harrowell} and meaningful quantities in
the $\alpha$-relaxation regime \cite{Gleim}.
This
means of extracting a critical density should actually yield the
mode-coupling density and not the glass transition density.  Thus, it
is not surprising that there is no thermodynamic signature in the
compressibility near $\rho_{c}=0.805$.  One can make a rough estimate
of the location of $\rho_{K}$ in the system of Santen and Krauth by
noting that a conservative guess for the ratio of mode-coupling to
Kauzmann densities from the $ \rho T^{-\frac{D}{12}} $ dependence in an ``equivalent''
soft-sphere system would imply that $\rho_{K}$ is about 10 percent
larger than $\rho_{c}$.  While this estimate is crude, {\em it implies
that exploration of the putative entropy crisis in the system studied
by Santen and Krauth is out of reach of both the SSMC and cluster
algorithm used by Santen and Krauth.}

To further explore the properties of this two dimensional hard-sphere
system, we address the nature of the phase space available to the
smaller discs.  While, at the highest densities we can probe, the
largest discs are completely trapped for long Monte Carlo times,
the smaller discs are diffusive, and do not even show a
significant cage effect.  In Fig. 5 we show the mean square
displacement of large and small discs, illustrating that part of the
system remains fluid even at the highest densities.  From the
standpoint of the standard definition of configurational entropy, it
would appear that the polydisperse hard-sphere system studied by
Santen and Krauth cannot experience a strict entropy crisis in
principle due to the disparity of disc sizes.  Recent work by Kumar {\em et.
al.} suggests that a reasonable correlation should exist between the density
of random close-packing and the Vogel density in a hard sphere system 
\cite{Kumar}. If we crudely interchange the Vogel density and the Kauzmann
density, then we expect that the ideal glass transition, if one occurs at all,
will occur close to the random close-packed density. If, as in the
polydisperse system of Santen and Krauth, the larger spheres freeze while the 
smaller spheres remain mobile, then the existence of the mobile, 
incompressible background fluid will always prevent an effective close-packing
of the larger spheres.
It is not clear what
signatures of an entropy crisis might exist in systems where a finite
fraction of particles may still undergo diffusive motion while the
remaining particles are frozen, but it would appear that the very concept
of an entropy crisis in such a system is perhaps ill-defined.  

Since the system studied by Santen and Krauth is both athermal and two
dimensional, we have also studied a three dimensional soft-sphere
version of this system.  In Fig. 6 we show the specific heat of a
small (N=34) version of the polydisperse three dimensional soft-sphere
system discussed in the introduction.  Interestingly, while the data
%
shows that the heat capacity calculated via fluctuations and by
derivatives of the average energy agree, there is a sharp
thermodynamic feature in the heat capacity at low temperatures.  One
would naively expect that this system cannot exhibit a state with
intermediate range order and therefore this peak must be related to a
glassy phenomenon, but this is not the case. Subtle features in the
radial distribution function as well as close inspection of instantaneous and
inherent structures make it clear that particles of similar size have
bunched together. In Fig. 7 we show
one such inherent configuration found via SSMC at low temperatures.

We conclude once again that the peak in the heat capacity is the
result of an exotic crystallization phenomena.  In this sense, our 
three dimensional thermalized version of the 
system studied by Santen and Krauth is similar to the binary
soft-sphere system studied by Grigera and Parisi.  For larger systems
the range of thermodynamic parameters for which non-local Monte Carlo
methods provide equilibrated results is not broad enough to
investigate deeply glassy states, while smaller systems show
thermodynamic signatures consistent with the formation of defective or
phase separated
crystals.  Interestingly, we could not find evidence of this type of
phase separation in small, two dimensional polydisperse hard or
soft-sphere systems, however we were still limited to densities and
temperatures that differ significantly from those at which an entropy crisis 
might be
expected.  

\vskip .2truein
\section{Conclusions}
\vskip .2truein

In this work we have taken a closer look at the recent use of
non-local Monte Carlo methods to investigate the nature of low
temperature glassy thermodynamics.  We have found that claims for and
against the existence of an entropy crisis that have been recently 
reported in the literature must be reconsidered.  The peak in the specific heat
found by Grigera and Parisi through the SMC approach can be
attributed to an energy gap between the manifold of defective
crystalline configurations and the phase separated ground state of the
system.  The use of SSMC (optimized SMC) performs at least as well as
the cluster algorithm of Santen and Krauth for the two dimensional
polydisperse hard-sphere system, but both are unable to probe this
system at high enough densities to render strong claims about the
behavior of the configurational entropy meaningful.  Furthermore,
since this system has a very wide range of particle radii, a
macroscopic portion of the system remains fluid at the highest
densities.  This fact further clouds the discussion of a possible
entropy crisis in this system.  It is not surprising that even
powerful Monte Carlo methods that make use of non-local moves still
cannot equilibrate systems under conditions where the number of
available configurations become sparse.  Interestingly, while the
Monte Carlo methods investigated here greatly increase the rate of
equilibration for temperatures (densities) near $T_{c}$ ($\rho_{c}$)
they are not efficient at lower temperatures (higher densities).  Completely 
polydisperse
systems are an exception to this finding, but these systems are
pathological in the sense that there is always significant phase space
for a finite fraction of the system to explore even when part of the
system is configurationally frozen.  In general, we have found that
one has to be careful to avoid the subtle phase separation and the corresponding
sharp features in thermodynamic quantities that may occur
in these systems.

While the findings presented here are somewhat negative, they do
provide important guidelines and warning signs for any computational
investigation of the entropy crisis in systems with realistic
potentials.  It would be interesting to perform non-local Monte Carlo
studies on small systems whose potentials are tailored to strongly
penalize structures with intermediate and long range order \cite{Leonardo}. 
It would also be interesting to re-investigate larger versions of the systems
studied here with the more powerful Wang-Landau approach \cite{WangWang2, Faller}. We are currently
pursuing these directions.

\begin{acknowledgements}
We gratefully acknowledge NSF support through grant \#CHE-0134969 and an NSF 
fellowship to Y.B.. D.R.R. is and Alfred P. Sloan Foundation Fellow and a 
Camille Dreyfus Teacher-Scholar.
\end{acknowledgements}

\pagebreak[4]
\begin{center}
\large{\bf{Figure Captions}}
\end{center}

Fig. 1:
(a) The heat capacity per particle
for a 34 particle binary soft-sphere system. The inset extends this calculation
to higher temperatures (lower $ \Gamma $), where a second smaller peak is evident.
$ C_V $ as calculated by fluctuations is plotted as (x) and finite difference
as (*), a convention which is maintained throughout the paper. Swap moves
were attempted on average every ten moves, and data was collected for between
$ 1\times 10^7 $and $1\times10^9 $
steps, with longer runs for lower temperatures.
(b) The heat capacity per particle for the same system and computational
details as in part (a), but with 60 particles. The heat capacity as calculated 
by fluctuations and derivatives is shown. The 34 particle peak is
included (solid line) to facilitate direct comparison.

Fig. 2:
Crystalline structures found by SMC. (a)
shows an inherent configuration of a 60 particle binary soft-
sphere system sampled at $\Gamma = 1.3 $, near the first heat capacity
peak and (b) presents an instantaneous structure sampled at $\Gamma = 1.8 $, near
the second peak.

Fig. 3:
The heat capacity per particle for the 70 particle binary soft-sphere
system. Equilibrated samples are obtained only up to $ \Gamma = 1.45 $. Swap moves
were attempted on average every ten moves, and data was collected for between $ 1\times 10^8 $ and $ 1 \times 10^9$
steps.

Fig. 4:
The converged inverse isothermal compressibility $\kappa^{-1}$ vs. $\rho$ for
the polydisperse hard-sphere system of Santen and Krauth. The system was
equilibrated by SSMC. Swap moves were attempted on average every ten moves and
data was collected for $1\times 10^9 $ steps, though such a long run was only necessary
for extremely high pressures.

Fig. 5:
Mean square displacement vs. MC time for two small and two large
particles chosen at random from the polydisperse hard-sphere system of Santen
and Krauth. The data is taken at $ P = 0.09 $, which corresponds to $\rho \approx
0.861 $. 

Fig. 6:
The heat capacity per particle for a 34 particle polydisperse
soft-sphere system. The peak arises due to a phase separation process
that induces intermediate range order in the system. Swap moves are attempted
on average every ten steps and data was collected for $ 1\times 10^7 $ to $ 6 \times 10^8 $ steps.

Fig. 7:
An inherent structure of the 34 particle polydisperse soft-sphere
system at $\Gamma = 10.0$. A cluster of large particles lies in the foreground,
and a cluster of small particles inside (not behind) the large cluster.
Similar clustering and the beginnings of phase separation
can be found in all low temperature structures of this system.

\clearpage
\pagebreak[4]

Fig. 1a
\begin{figure}[tbh]
\begin{center}
\includegraphics[width=\linewidth]{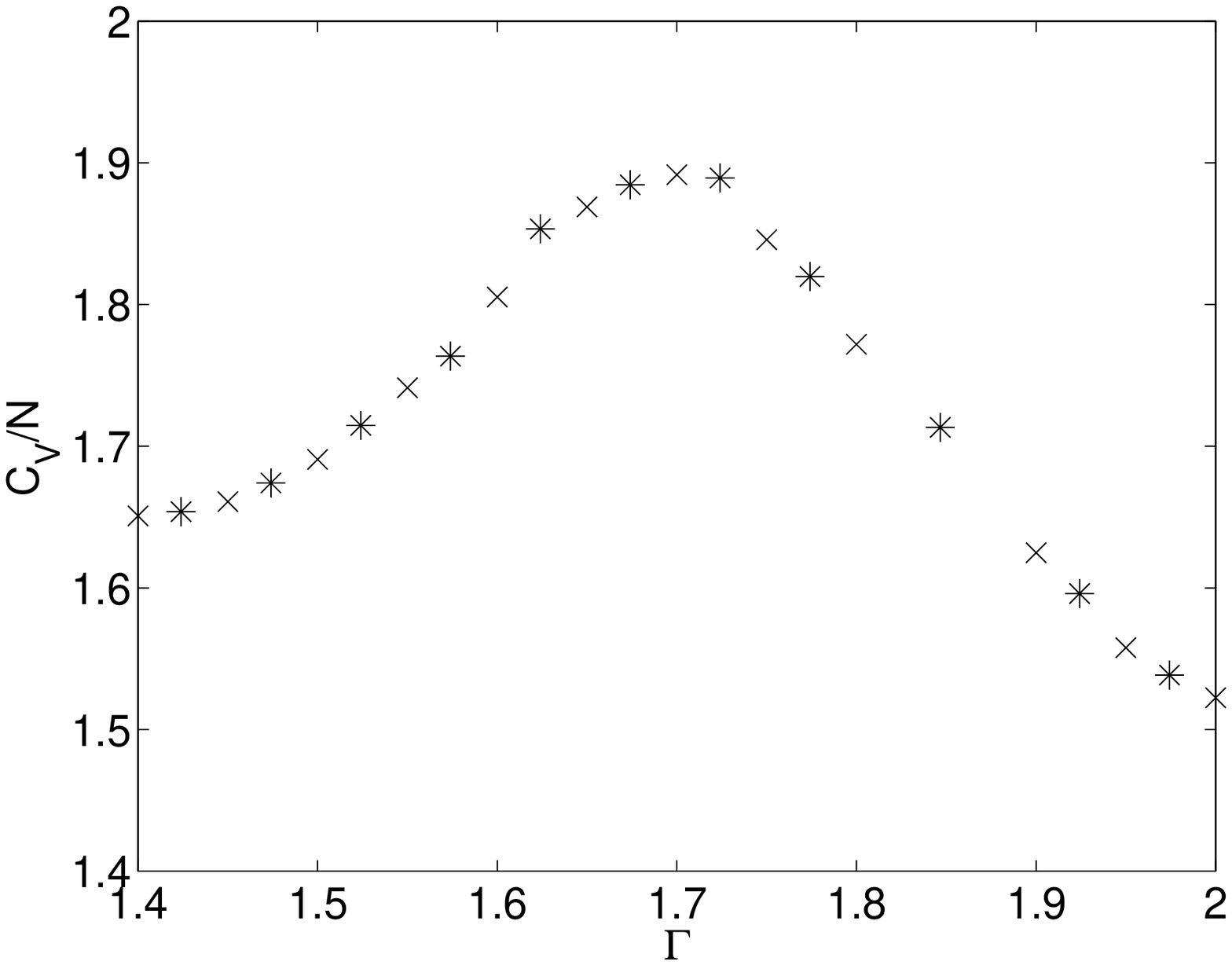}
\end{center}
\end{figure}
\begin{figure}[tbh]
\begin{center}
\vspace{-3.3in}
\includegraphics[width=0.4\linewidth]{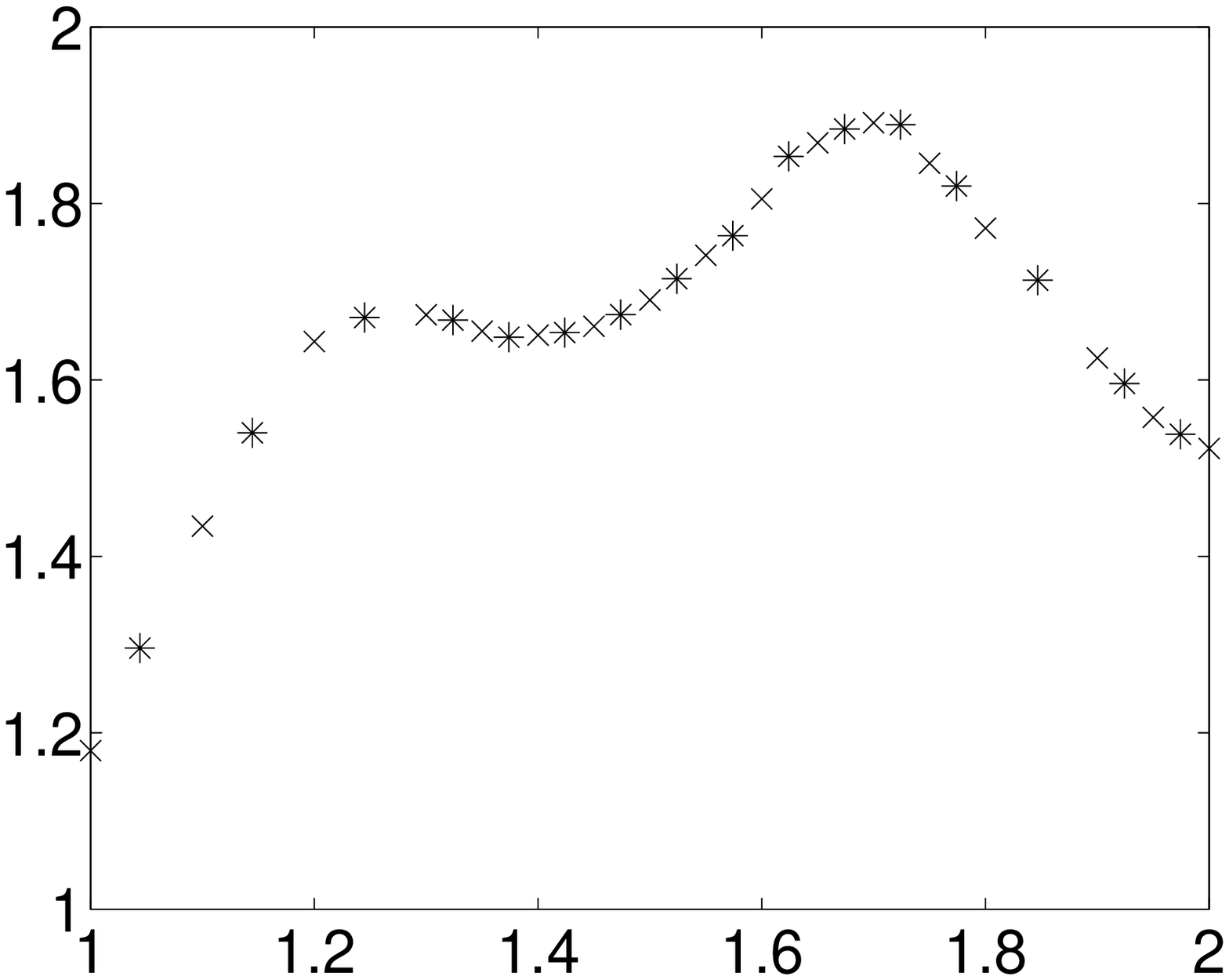}
\end{center}
\end{figure}

\clearpage
\pagebreak[4]

Fig. 1b

\begin{figure}[tbh]
\begin{center}
\includegraphics[width=\linewidth]{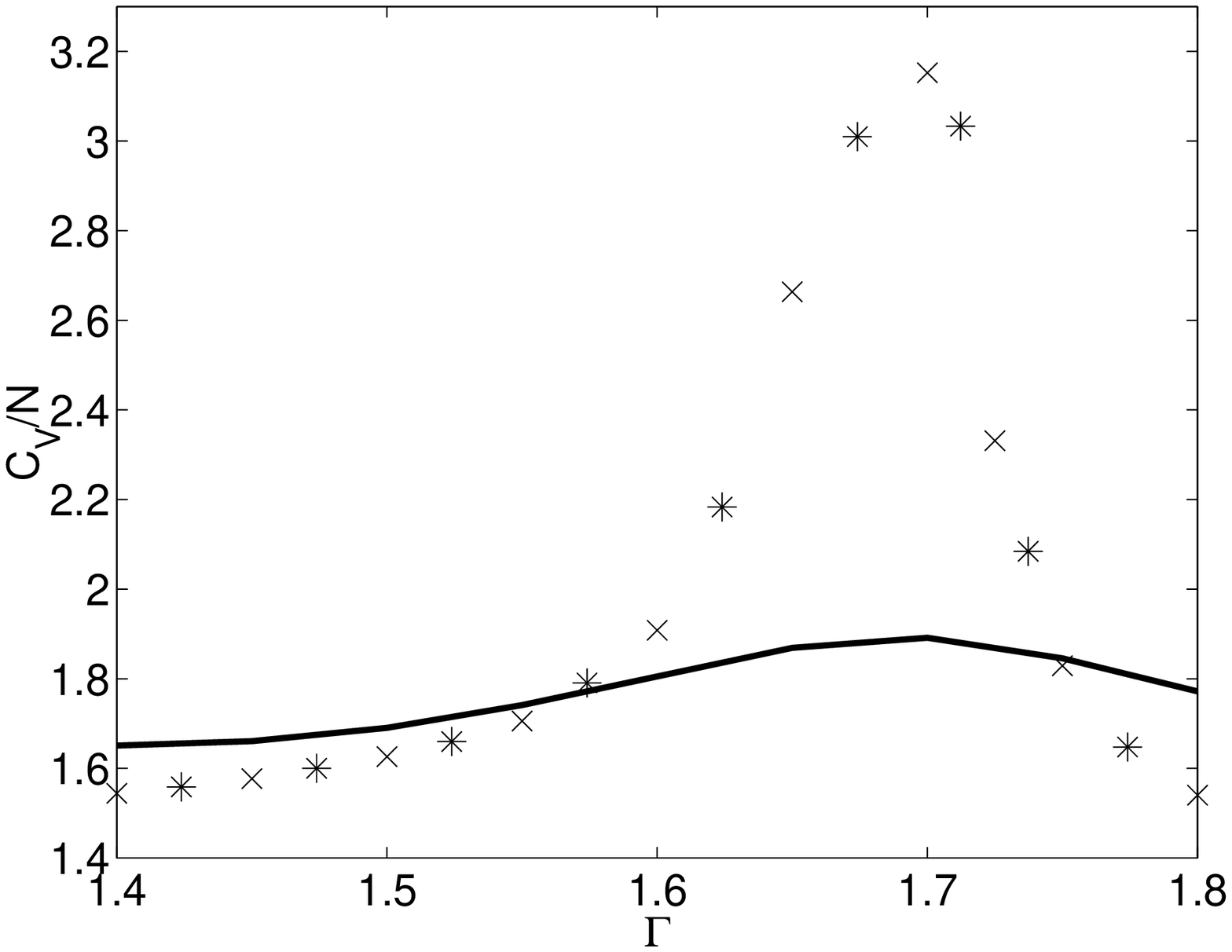}
\end{center}
\end{figure}

\clearpage
\pagebreak[4]

Fig. 2
\begin{figure}[tbh]
(a) \hspace{290pt}(b)
\vspace{20pt}\includegraphics[width=0.45\linewidth]{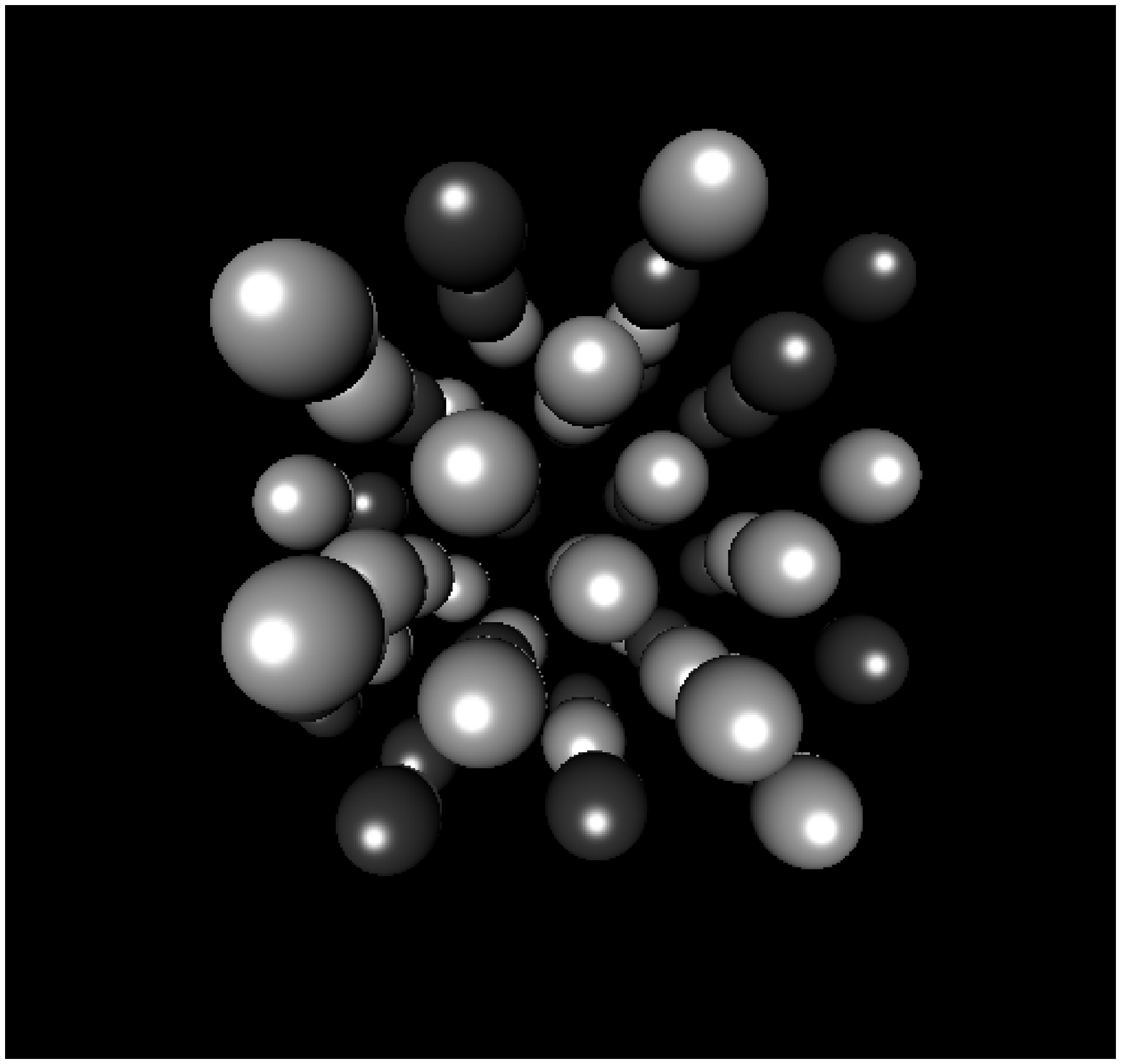}\hfill
\includegraphics[width=0.45\linewidth]{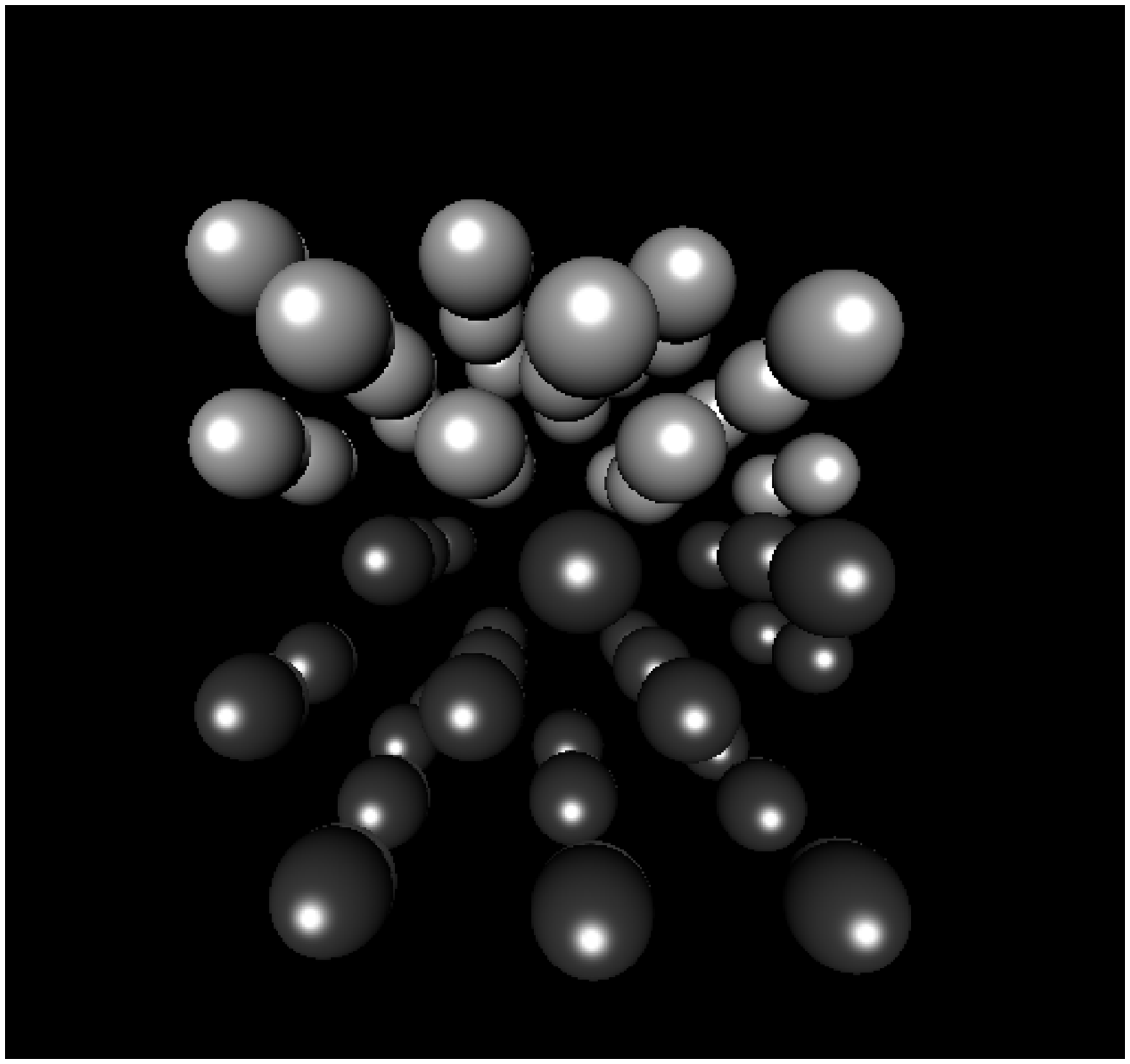}
\end{figure}
\clearpage
\pagebreak[4]

Fig. 3
\begin{figure}[tbh]
\begin{center}
\includegraphics[width=1.0\linewidth]{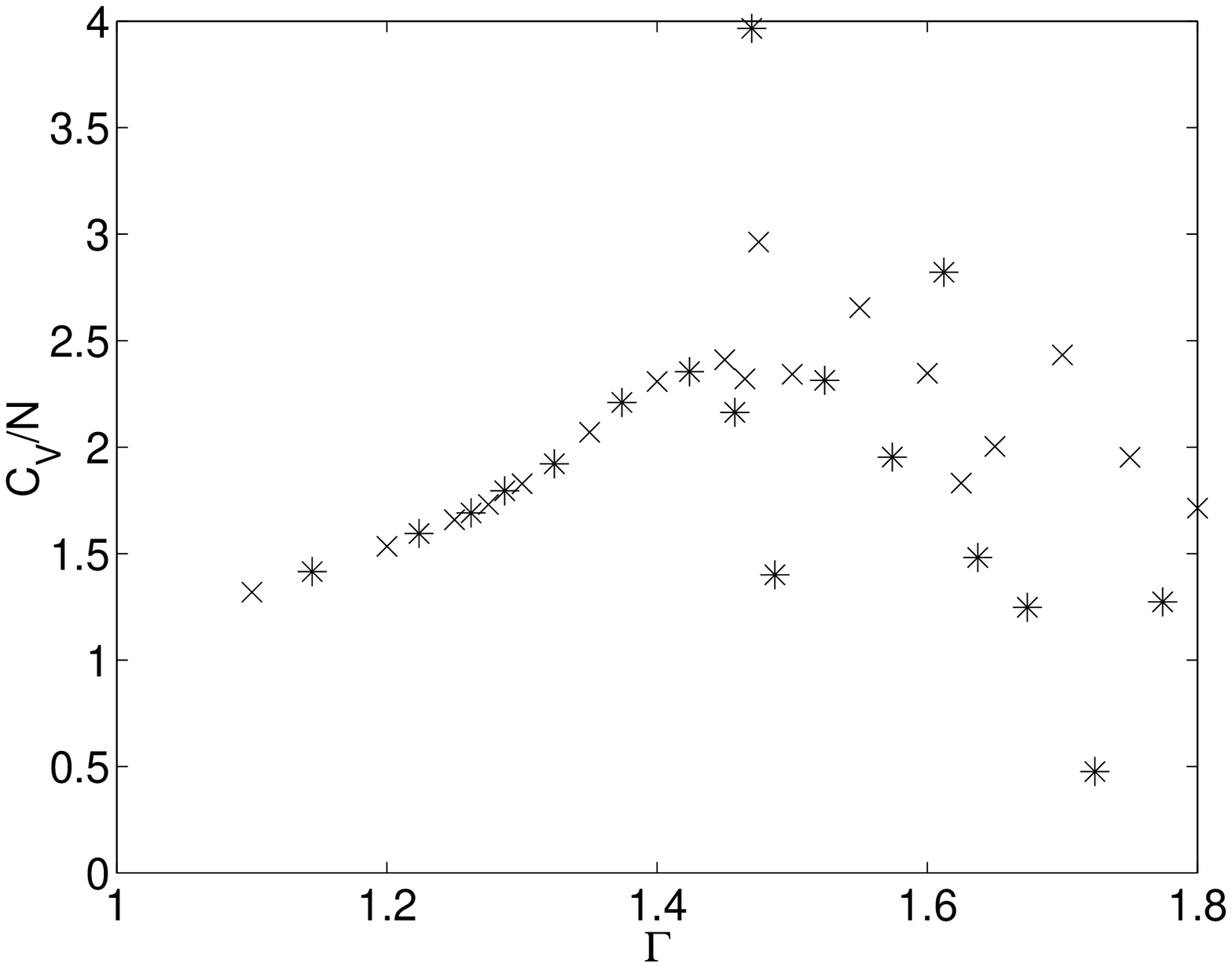}
\end{center}
\end{figure}
\clearpage
\pagebreak[4]

Fig. 4
\begin{figure}[tbh]
\begin{center}
\includegraphics[width=1.0\linewidth]{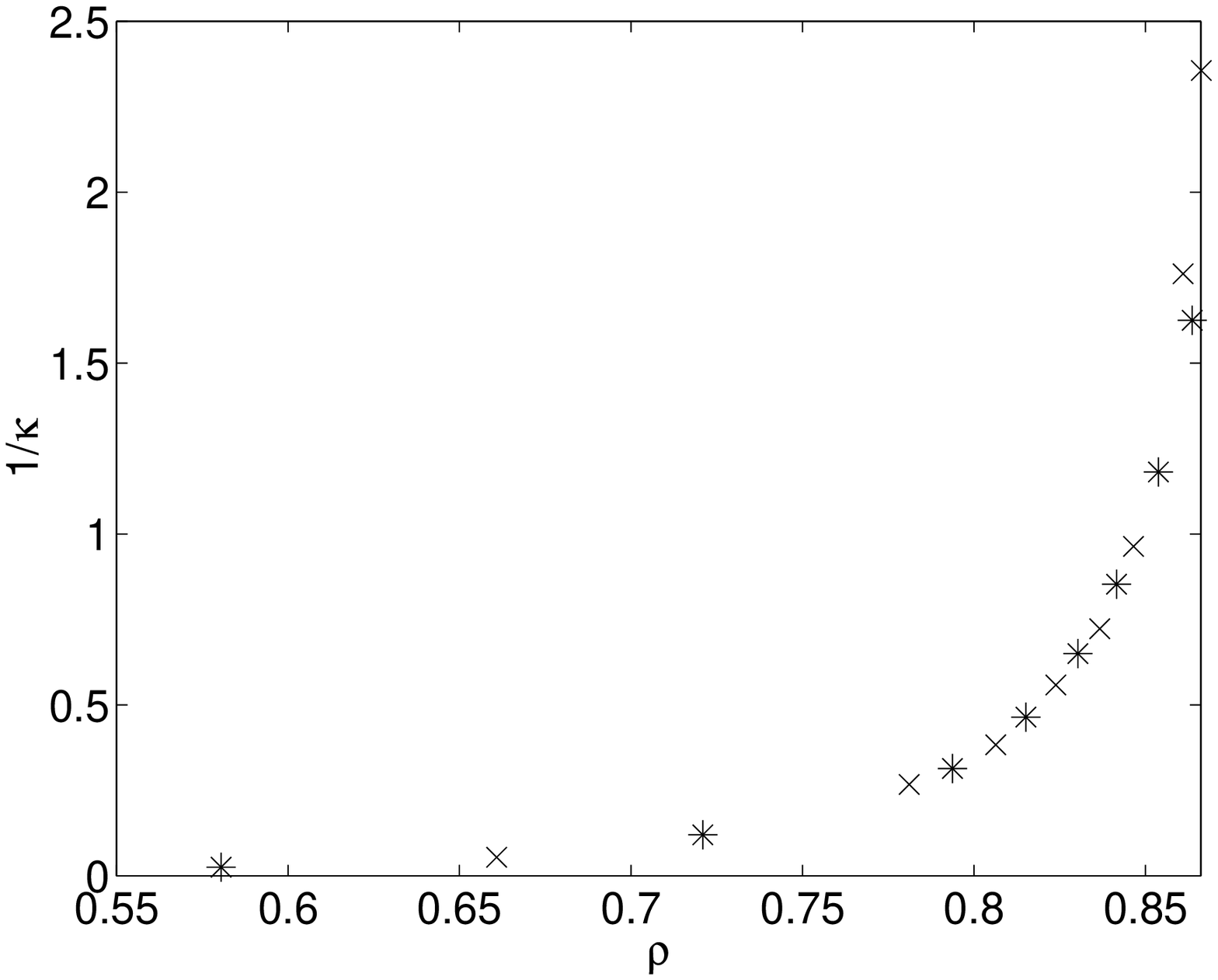}
\end{center}
\end{figure}
\clearpage
\pagebreak[4]

Fig. 5
\begin{figure}[tbh]
\begin{center}
\includegraphics[width=1.0\linewidth]{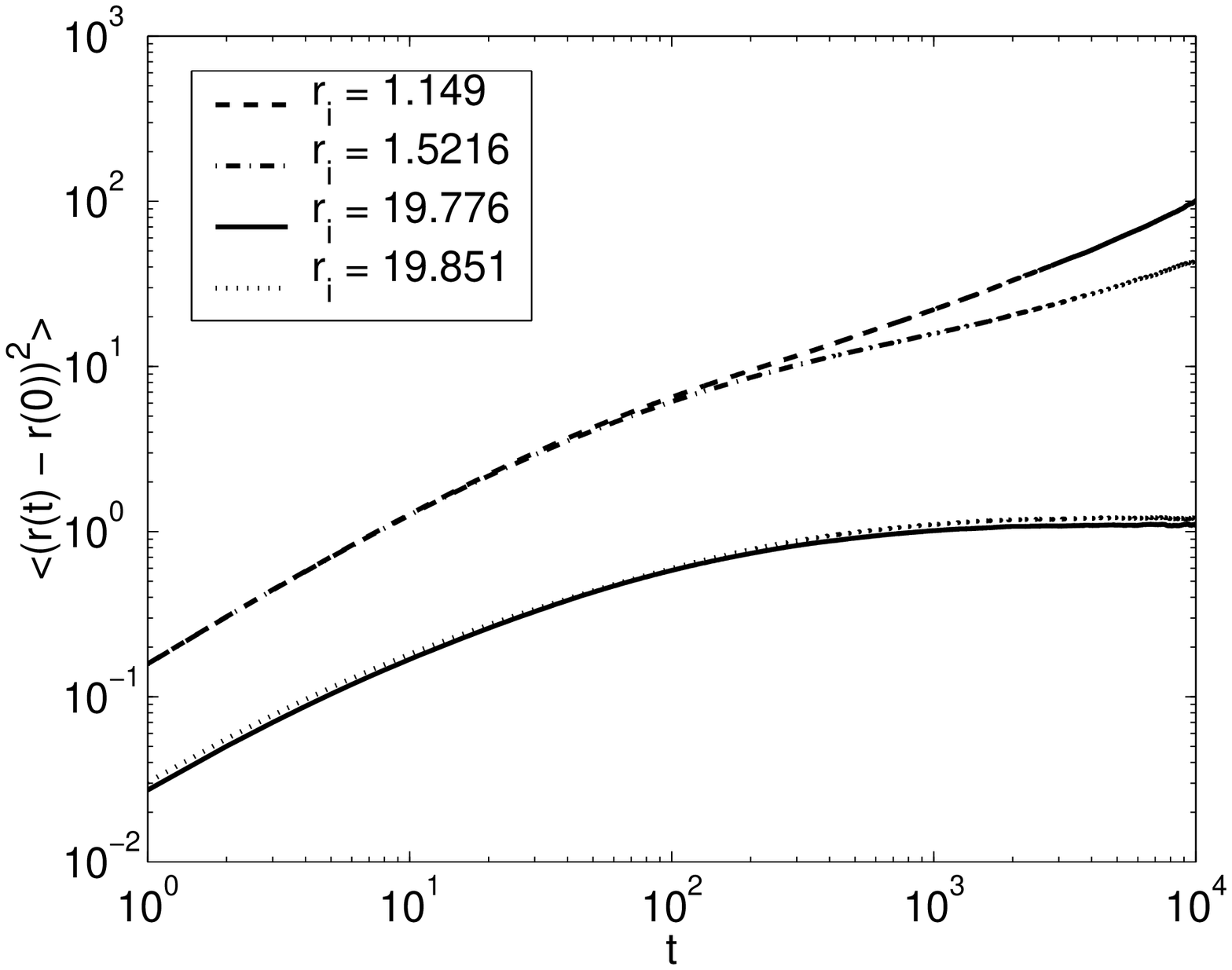}
\end{center}
\end{figure}
\clearpage
\pagebreak[4]

Fig. 6
\begin{figure}[tbh]
\begin{center}
\includegraphics[width=1.0\linewidth]{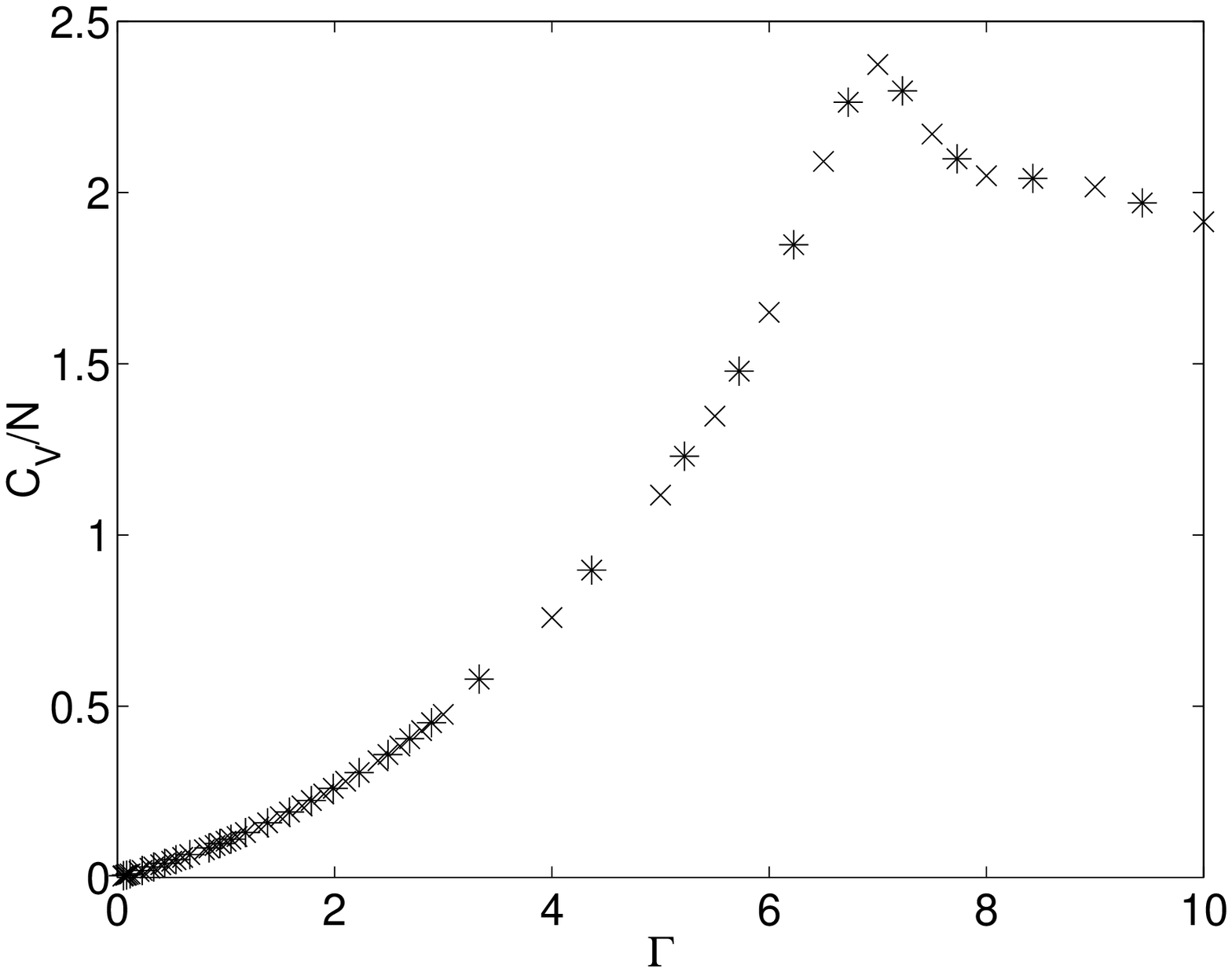}
\end{center}
\end{figure}

\clearpage
\pagebreak[4]
\vspace{2in}
Fig. 7
\begin{figure}[tbh]
\begin{center}
\includegraphics[width=1.0\linewidth]{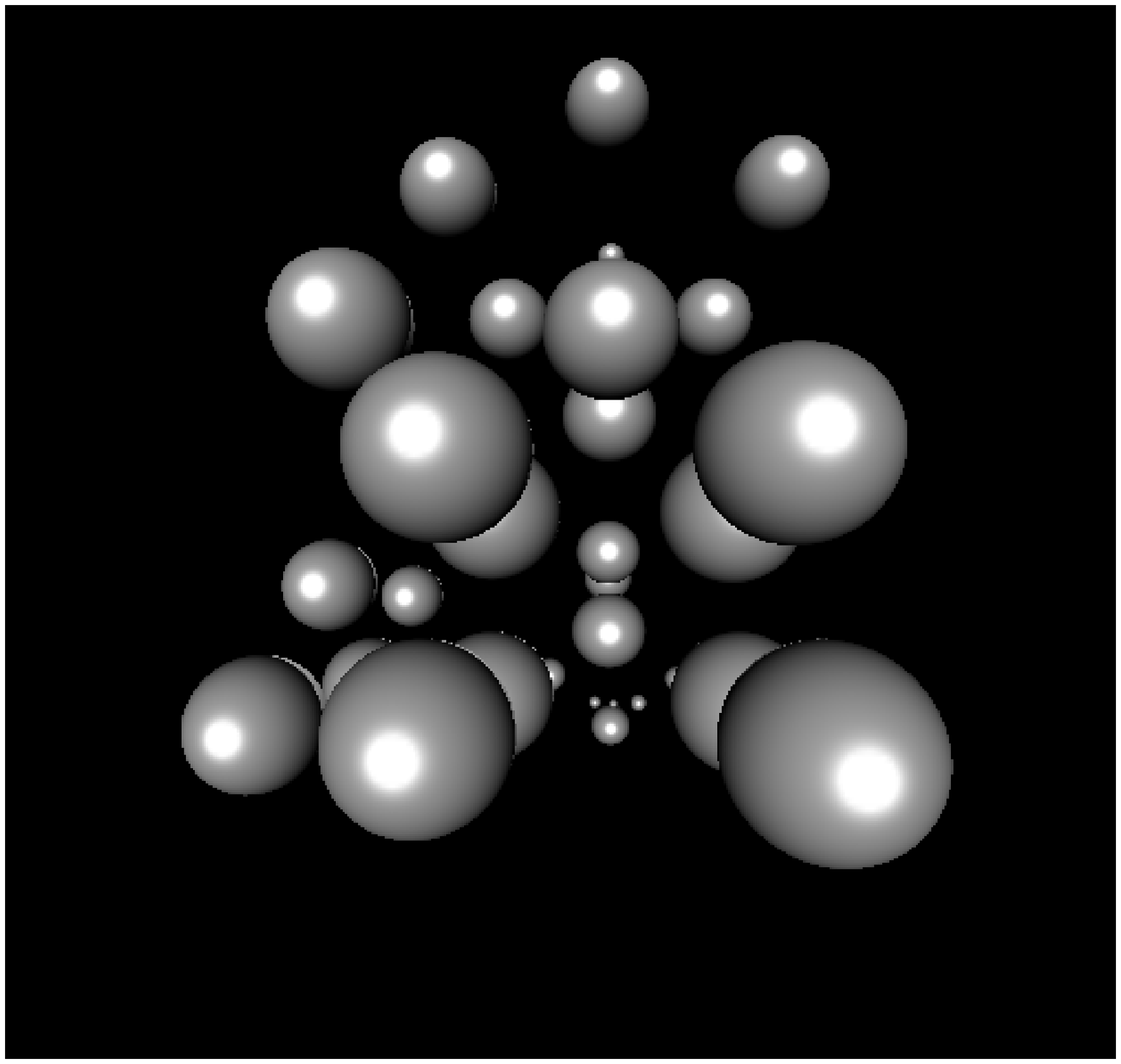}
\end{center}
\end{figure}
\clearpage
\pagebreak[4]

\end{document}